\documentstyle[12pt]{article}

\advance\textheight 1cm
\advance\topmargin -0.5cm
\begin{document}

\title{Unitarity Restoration in the Presence\\ of Closed Timelike Curves}

\author{Arlen Anderson\thanks{arley@physics.unc.edu}\\
Isaac Newton Institute\\
20 Clarkson Road\\
Cambridge CB3 0EH, England\\
and\\
Blackett Laboratory\\
Imperial College\\
Prince Consort Rd.\\
London SW7 2BZ, England\\
and\\
Dept. of Physics and Astronomy\thanks{Present Address}\\
UNC-Chapel Hill\\
Chapel Hill NC 27599-3255}
\date{May 25, 1994, revised Nov. 10, 1994}
\maketitle
\vspace{-14cm}
\hfill Imperial/TP/93-94/41

\hfill gr-qc/9405058
\vspace{12cm}

\begin{abstract}
A proposal is made for a mathematically unambiguous treatment of
evolution in the presence of closed timelike curves.  In
constrast to other proposals for handling the naively nonunitary
evolution that is often present in such situations, this proposal
is causal, linear in the initial density matrix and preserves probability.
It provides a physically reasonable interpretation of invertible nonunitary
evolution by redefining the final Hilbert space so that the evolution
is unitary or equivalently by removing the nonunitary part of the
evolution operator using a polar decomposition.
\end{abstract}

\newpage

Evolution in spacetimes containing closed timelike curves has drawn much
attention[1-8] recently following the work of Morris, Thorne and Yurtsever
on the possibility of time machines\cite{MoT}.  Such a spacetime is
not globally hyperbolic and therefore does not admit a foliation by
spacelike hypersurfaces. Standard quantum field theory techniques cannot be
applied. Nevertheless it has been proven that free field evolution on
examples of such backgrounds is consistent and unitary while in the
case of interacting fields, the evolution is apparently nonunitary[5-7].
The concept of nonunitary evolution is sufficiently disturbing
that one might be tempted to dismiss these results out of hand.
This would be a mistake as a closer study is repaid by a clearer
understanding of {\it unitary} evolution.

There have been previous considerations of nonunitary evolution and proposals
for how to interpret it.  Jacobson argued that causality
implies unitarity, and that there are ambiguities in computing
expectation values for observables localized in regions spacelike separated
from a region of nonunitary evolution\cite{Jac}.  The possibility
was raised that a nonunitary Heisenberg formulation could be causal
and unambiguous.

An unambiguous but acausal path integral interpretation
based on the decoherence functional formalism was proposed by Friedman,
Papastamatiou and Simon\cite{FPS}. This was elaborated thoroughly by
Hartle\cite{Har}.  Both find that the presence of closed
timelike curves in the future acausally affect the probabilities
for observations in the present.  In addition, the decoherence
functional is nonlinear in the initial density matrix, and this
represents a nonlinear (but unobservable) modification of standard
quantum mechanics.

In this paper, after reviewing the discussions of Jacobson and Hartle,
a proposal is made for an unambiguous and causal treatment of
(invertible) nonunitary evolution which is linear in the initial
density matrix.  This proposal is given both in the language of states
and as a decoherence functional.  The resulting description in fact shows
that the evolution is unitary when appropriately handled.

The essence of the proposal
follows from the observation that in [5-7] the evolution $X$
is identified as nonunitary because $X X^{\dagger}\ne 1$, but
this is not sufficient to conclude nonunitarity in
the sense of loss of probability \cite{And}.
A unitary operator $U$ is defined\cite{ReS} to be a linear operator from one
Hilbert space ${\cal H}_a$ onto a second ${\cal H}_b$ such that
\begin{equation}
\label{def}
\langle U\psi|U\phi\rangle_{{\cal H}_b}=\langle \psi|\phi\rangle_{{\cal H}_a},
\end{equation}
for all $\psi,\ \phi\in {\cal H}_a$.  Because all Hilbert spaces with a
countable basis are isomorphic\cite{ReS}, it is conventional to write
inner products
formally without stating their measure densities explicitly.  Additionally,
in standard quantum mechanical examples, the Hilbert space obtained
after evolution is the same as the initial one and the inner products
are the same.  This combination of facts makes it easy to overlook
that the final inner product and Hilbert space must be specified
when defining the adjoint.  $\langle X\psi|X\phi\rangle_{{\cal H}_a}$ and
$\langle X\psi|X\phi\rangle_{{\cal H}_b}$ are not the same unless
${\cal H}_a={\cal H}_b$.  Thus, $X$ may not be unitary
($X^\dagger X\ne 1$) when the final Hilbert space is taken to be the same
as the initial one, but nevertheless $X$ is
unitary in the sense of (\ref{def}) for a suitably chosen final Hilbert
space.   The proposal for handling (invertible) naively nonunitary $X$ is to
work with the final Hilbert space where $X$ is unitary.
This is equivalent to remaining in the original Hilbert space ${\cal H}_a$ but
removing the nonunitary part of $X$ through
a polar decomposition $X=RU_X$, where $U_X$ is unitary
$U_X U_X^{\dagger_a}=1$ and $R^2=X X^{\dagger_a}$ is self-adjoint in
${\cal H}_a$. Then, $U_X$ is the manifestly unitary evolution operator in
${\cal H}_a$. This will be discussed in detail below.

One's initial concern with nonunitary evolution is that probability must
not be conserved, and that therefore one cannot compute expectation values.
A proposal to get around this is to simply renormalize expectation values
by the (changing) non-unit norm of the state in which the expectation is
taken
\begin{equation}
\langle P_\alpha \rangle= {\langle \psi | P_\alpha | \psi \rangle \over
\langle \psi | \psi \rangle}.
\end{equation}
This gives a well-defined expectation value even when the norm of
$\langle \psi | \psi \rangle$ is not unity.  The difficulty with this
proposal, as shown by Jacobson\cite{Jac}, is that it can give ambiguous
results.

Consider a measurement of $P_\alpha$ made in a region $A$ well-separated
in a spacelike direction from a region of nonunitary evolution, for
instance a region containing closed timelike curves (CTCs).  The expectation
value of $P_\alpha$ can be computed on either a spacelike hypersurface
$\sigma_-$ through $A$ that passes before the CTC region or on a spacelike
hypersurface $\sigma_+$ through $A$ that passes after the CTC region.
The wavefunction $|\psi\rangle$ is different on these two hypersurfaces and
is related by the nonunitary evolution $X_N$ through the CTC region
\begin{equation}
|\psi\rangle_+= X_N|\psi\rangle_-
\end{equation}
where the subscript labels the hypersurface on which the wavefunction is
evaluated.  The expectation value of $P_\alpha$ on $\sigma_-$ is then
\begin{equation}
\langle P_\alpha \rangle_- ={\phantom{\langle}_- \langle \psi | P_\alpha
| \psi \rangle_- \over
\phantom{\langle}_-\langle \psi | \psi \rangle_-},
\end{equation}
while on $\sigma_+$, it is
\begin{equation}
\langle P_\alpha \rangle_+ ={\phantom{\langle}_- \langle \psi |X_N^\dagger
P_\alpha X_N| \psi \rangle_- \over
\phantom{\langle}_-\langle \psi |X_N^\dagger X_N| \psi \rangle_-}.
\end{equation}
Since the observable $P_\alpha$ is measured in the same region on both slices,
it has not changed under evolution, that is, $X_N^{-1} P_\alpha X_N=
P_\alpha$.  This implies that
\begin{equation}
\langle P_\alpha \rangle_+ ={\phantom{\langle}_- \langle \psi |X_N^\dagger
X_N P_\alpha | \psi \rangle_- \over
\phantom{\langle}_-\langle \psi |X_N^\dagger X_N| \psi \rangle_-},
\end{equation}
which is not equal to $\langle P_\alpha \rangle_-$ unless $X_N^\dagger
X_N=1$.  Thus,  for this proposal,
nonunitary evolution leads to ambiguous predictions for expectation
values of observables even in regions spacelike separated from the region
of nonunitary evolution.

A different proposal\cite{Har} using a decoherence functional in Gell-Mann
and Hartle's generalized quantum mechanics\cite{GMH,Har2} succeeds in
avoiding this ambi\-guity. The decoherence functional measures the
interference between histories characterized by strings of projections. The
projections serve to divide the set of fine-grained histories into
coarser-grained classes. The decoherence functional can be computed as a
closed-time-path functional integral over histories which start from an
initial density matrix, pass forward through one sequence of projections
up to a final surface, and then back through a different set. With no final
condition, one traces over a final surface. In the Heisenberg picture of the
unitary case, projection operators on a spacelike slice $\sigma_k$ can be
expressed in terms of the unitary evolution $U(\sigma_k,\sigma_a)$ of
projection operators defined on an initial slice $\sigma_i$
\begin{equation}
P_{\alpha_k}^{(k)}(\sigma_k)= U^{-1}(\sigma_k,\sigma_i) P_{\alpha_k}^{(k)}
U(\sigma_k,\sigma_i).
\end{equation}
An overall unitary operator evolving from the initial surface $\sigma_i$
to the final surface $\sigma_f$ cancels by cyclicity of the trace and the
decoherence functional is left as a trace over the initial hypersurface
\begin{equation}
D(\alpha',\alpha)={ \rm Tr}\biggl( P_{\alpha'_n}^{(n)}(\sigma_n)\cdots
P_{\alpha'_1}^{(1)}(\sigma_1) \rho_i
P_{\alpha_1}^{(1)\dagger}(\sigma_1)\cdots
P_{\alpha_n}^{(n)\dagger}(\sigma_n) \biggr).
\end{equation}

In the non-unitary case discussed above, there is a region bounded by
$\sigma_+$ and $\sigma_-$ containing a
non-unitary evolution $X_N$. This evolution must be included in the
decoherence functional: $X_N$ is inserted in the sequence of projections
where it takes place. Projections before $\sigma_-$ are expressed
in the Heisenberg picture in terms of the initial hypersurface $\sigma_i$,
and those after $\sigma_+$ in terms of projections on the final hypersurface
$\sigma_f$
\begin{eqnarray}
\label{projs}
P_{\alpha_k}^{(k)}(\sigma_k)&=& U^{-1}(\sigma_k,\sigma_i) P_{\alpha_k}^{(k)}
U(\sigma_k,\sigma_i), \quad \sigma_k < \sigma_- \\
P_{\alpha_k}^{(k)}(\sigma_k)&=& U^{-1}(\sigma_k,\sigma_f)P_{\alpha_k}^{(k)}
U(\sigma_k,\sigma_f), \quad \sigma_k > \sigma_+. \nonumber
\end{eqnarray}
To make the transition between the initial and final surfaces, one uses the
full evolution
\begin{equation}
X=U(\sigma_f,\sigma_+)X_N U(\sigma_-,\sigma_i).
\end{equation}
Additionally, the decoherence functional must be
renormalized to account for the nonunitary evolution of the initial
density matrix in the absence of coarse-graining.  The proposed
decoherence functional in
the case of one projection made before the CTC region is\cite{Har}
\begin{equation}
D_-(\alpha',\alpha)={{ \rm Tr}\biggl( XP_{\alpha'}^{(1)}(\sigma_1) \rho_i
P_{\alpha}^{(1)\dagger}(\sigma_1)X^\dagger \biggr) \over
{ \rm Tr}\biggl( X \rho_i X^\dagger \biggr)} .
\end{equation}
For the case of one projection made after the CTC region, it would be
\begin{equation}
\label{ndf}
D_+(\alpha',\alpha)={{ \rm Tr}\biggl( P_{\alpha'}^{(1)}(\sigma_1)
X\rho_i X^\dagger P_{\alpha}^{(1)\dagger}(\sigma_1) \biggr) \over
{ \rm Tr}\biggl( X \rho_i X^\dagger \biggr)} .
\end{equation}
Now, if the projection is made in a region spacelike separated from the
CTC region, then as above $P_\alpha$ commutes with $X$ and one sees that
$D_-(\alpha',\alpha)=D_+(\alpha',\alpha)$.  There is no ambiguity.  It has
disappeared because both computations are aware of the full future
history, and in particular of the presence of the CTC region.

This proposal has a curious property however.  The result of computing
the decoherence functional is different in the presence of $X$ than it is
in its absence.  If we are living in the past of a CTC region, we might
naively compute with a decoherence function which did not include $X$.
Our predictions would be different than they would be if we included a
nonunitary $X$.   If the correct physical description involved $X$, the
difference between physical results and our no-$X$ predictions would in
principle be susceptible to experimental
detection.  With this decoherence functional, the future existence of a
CTC region acausally affects the accuracy of predictions today.  This is
a strange result, but very important if true.

A second unusual feature of this decoherence functional (and indeed the
earlier proposal) is that it is nonlinear in the initial density
matrix.  That is, because the trace of the initial density matrix
appears in the denominator, one cannot add different initial
density matrices.  This is therefore a nonlinear modification of conventional
quantum mechanics.  Arguably in a closed-system formulation of the
wavefunction of the universe, we would not be sensitive to this in any
event as we only see linearity in the superpositions of subsystems, each
of which are part of the full initial density matrix.

There is however an alternative.  A new proposal leads to both
unambiguous results and is causal and linear.  This proposal is
based on the observation that a linear invertible transformation
$X$, such that $X X^{\dagger_a}\ne 1$, is not a unitary transformation
from the Hilbert space ${\cal H}_a$ to itself, but it is a unitary
transformation to a new Hilbert space ${\cal H}_b$\cite{And}.  The
states obtained by unitary evolution with $X$ are elements of
${\cal H}_b$, and they must be transformed back to ${\cal H}_a$ if
one is to compare with the physical states defined there.  After this
is done, one finds that these evolved states
are simply those obtained by evolving with the unitary part $U_X$ of a
polar decomposition of $X=RU_X$ [where $R=(X X^{\dagger_a})^{1/2}$ is
self-adjoint in ${\cal H}_a$].  Unitarity is restored by removing the
nonunitary part of $X$. This proposal can be implemented both in the
language of states and expectation values and in the decoherence functional.

The manipulations needed to elaborate the proposal can all
be performed symbolically, but for clarity and familiarity,
the case of quantum mechanics is treated explicitly first.
In the language of states, one has to be careful about the form of the
inner product when it is computed on different hypersurfaces after
evolution.  The inner product on an initial hypersurface $\sigma_i$ for
states in the Hilbert space ${\cal H}_a$ is
\begin{eqnarray}
\label{ip}
\phantom{\langle}_{(\mu_a)} \langle \phi,i | \psi,i \rangle &=&
\int_{\sigma_i} dx \phi^* \mu_a \psi \\
&=& \phantom{\langle}_{(1)} \langle \phi,i |\mu_a| \psi,i \rangle, \nonumber
\end{eqnarray}
where the (possibly operator-valued) measure density in the inner product
is pre-subscripted parenthetically. (The measure density is put in front
to emphasize that it is more naturally associated with the bra-vectors than
the kets.)  Let $|\tilde\psi,f\rangle=X|\psi,i\rangle$ be the state on
the final hypersurface $\sigma_f$ obtained by evolving an initial state
on $\sigma_i$ with $X$---the tilde indicates that this is a state in
the Hilbert space ${\cal H}_b$.  Requiring that the value of the
inner product be preserved when evaluated on the hypersurface $\sigma_f$,
one has
\begin{eqnarray}
\label{pres}
\phantom{\langle}_{(\mu_a)} \langle \phi,i | \psi,i \rangle&=&
\int_{\sigma_i} dx (X^{-1}\tilde\phi)^* \mu_a X^{-1}\tilde\psi \nonumber\\
&=& \int_{\sigma_f} dx \tilde\phi^* X^{-1\,\dagger_1}\mu_a X^{-1}\tilde\psi
\nonumber\\
&=& \int_{\sigma_f} dx \tilde\phi^* \mu_a X^{-1\,\dagger_a}X^{-1}\tilde\psi \\
&=& \int_{\sigma_f} dx \tilde\phi^* \mu_b \tilde\psi
= \phantom{\langle}_{(\mu_b)} \langle \tilde\phi,f | \tilde\psi,f \rangle,
\nonumber
\end{eqnarray}
The induced measure density of the unitarily equivalent inner product is
\begin{equation}
\label{tmd}
\mu_b=(X^{-1})^{\dagger_1} \mu_a X^{-1}=\mu_a (X X^{\dagger_a})^{-1}.
\end{equation}
The notation $\dagger_1$ indicates that the adjoint is computed in the trivial
measure density $\mu=1$. Note that this adjoint is related to
the adjoint in the measure density $\mu_a$ by
\begin{equation}
X^{\dagger_1} \mu_a=\mu_a X^{\dagger_a}.
\end{equation}
One sees that to preserve the inner product the measure density changes
during the evolution,  acquiring an inhomogeneous factor
$(X X^{\dagger_a})^{-1}$ measuring the nonunitarity of $X$.

A side remark may prevent potential confusion.  Consider
the (nonrelativistic) quantum mechanics of a particle propagating in a
time-dependent curved background.  The measure density $\mu_a$ is
the square-root of the determinant of the metric and therefore is
time-dependent.  Nevertheless it is specified once for the whole spacetime
and the form of the inner product never changes; on each hypersurface
where one computes the inner product, one evaluates
$\mu_a$ at that instant of time.   Here, the transformation to
$\mu_b$ reflecting the nonunitarity of $X$ is on top of any
time-dependence $\mu_a$ may already have.  It is a genuine
change to a different Hilbert space.

Having found the Hilbert space in which $X$ is unitary, we must
face the fact that it is not the Hilbert space we naively expected.
Our usual association between mathematical and physical
quantities is made with the presumption of a particular
Hilbert space.  We know what physics means in the usual
Hilbert space, and we would be surprised if it made a difference to be
told that there had been ``nonunitary" evolution in our past and that
the Hilbert space at present is not the one we naively thought!  Fortunately,
we needn't worry:  the choice of measure density/Hilbert space
is on the same par as a choice of coordinates; you can
freely change it when talking about physical quantities.  We just need
to find the transformation from $\mu_b$ to $\mu_a$.

Let $R=(X X^{\dagger_a})^{1/2}$. Then $R=R^{\dagger_a}$ is
self-adjoint in ${\cal H}_a$. Returning to the computation in
Eq. (\ref{pres}), one sees that defining
\begin{equation}
\label{trans}
|\psi,f\rangle=R^{-1}|\tilde\psi,f\rangle,
\end{equation}
resets the measure to $\mu_a$,
\begin{equation}
 \int_{\sigma_f} dx \tilde\phi^* \mu_a R^{-2}\tilde\psi=
 \int_{\sigma_f} dx \phi^* \mu_a \psi.
\end{equation}
But this means that the evolved state in ${\cal H}_a$ is
\begin{equation}
|\psi,f\rangle=R^{-1}X|\psi,i\rangle=U_X|\psi,i\rangle.
\end{equation}
It is easy to check that $U_X$ is unitary in ${\cal H}_a$
($U_X U_X^{\dagger_a}= U_X^{\dagger_a} U_X=1$)! This, $X=RU_X$, is the polar
decomposition of $X$ in ${\cal H}_a$.  The restoration of unitarity
is explicit.

Under the transformation (\ref{trans}) from ${\cal H}_b$ to ${\cal H}_a$
on the final surface,
the self-adjoint observable $\tilde P_\alpha=\tilde P_\alpha^{\dagger_b}$ in
${\cal H}_b$ becomes
\begin{equation}
\label{optrans}
P_\alpha=R^{-1} \tilde P_\alpha R,
\end{equation}
which is self-adjoint in ${\cal H}_a$.   With this in mind, one can turn
to consider ``non-unitary" evolution in the Heisenberg picture.
The naive Heisenberg evolution of
the operator $P_\alpha$ is $X^{-1} P_\alpha X$.  Unfortunately this cannot be
correct because $X^{-1} P_\alpha X$ is not self-adjoint in ${\cal H}_a$
(that is, on the initial hypersurface).  This should not be surprising.

Heisenberg evolution works by pulling back operators from the final surface
to the initial surface.  If $\tilde P_\alpha$ is self-adjoint in ${\cal H}_b$
on the final surface, then it is easy to confirm that its pull-back
is self-adjoint on the initial surface in ${\cal H}_a$
\begin{eqnarray}
\label{sadja}
X^{-1}\tilde P_\alpha X= X^{-1}\tilde P_\alpha^{\dagger_b} X &=&
X^{-1}\mu_{b}^{-1}\tilde P_\alpha^{\dagger_1}\mu_{b} X \\
&=& \mu_a^{-1} X^{\dagger_1 }\tilde P_\alpha^{\dagger_1} (X^{-1})^{\dagger_1}
\mu_a \nonumber \\
&=& (X^{-1} \tilde P_\alpha X)^{\dagger_a}. \nonumber
\end{eqnarray}
Using Eq. (\ref{optrans}), this implies the Heisenberg evolution of $P_\alpha$
is
\begin{equation}
\label{Hop}
X^{-1}\tilde P_\alpha X= U_X^{-1} P_\alpha U_X.
\end{equation}
Heisenberg evolution is accomplished using the unitary part of the polar
decomposition of $X$.

Sandwiching this Heisenberg operator between initial states, one
sees that one has what one wants--the expectation of $P_\alpha$ in
${\cal H}_a$ on the final surface,
\begin{equation}
\phantom{\langle}_{(\mu_a)}\langle \psi,i|  U_X^{-1} P_\alpha U_X | \psi,i
\rangle= \phantom{\langle}_{(\mu_a)}\langle \psi,f|  P_\alpha | \psi,f \rangle.
\end{equation}
 From another perspective this result is natural:
in the Heisenberg picture, one does not have the freedom to change
the Hilbert space, so to restore unitarity one must change one's
identification of the evolved observable.

One might be concerned that the nonlinear operation involved in removing
the nonunitary part would prevent one from composing $X$ with
a further unitary operator, say $U_Y$.  Let $Z=U_Y X$ be
such a composition,  then
\begin{eqnarray}
U_Z&=&(Z Z^{\dagger_a})^{-1/2} Z \nonumber \\
&=& (U_Y R^{2} U_Y^{-1})^{-1/2} U_Y R U_X \\
&=& U_Y U_X. \nonumber
\end{eqnarray}
Here one can use $U_Y R^2 U_Y^{-1} =( U_Y R U_Y^{-1})^2$ before
taking the square root, but it is generally true that
$f( U_Y R U_Y^{-1})=U_Y f(R) U_Y^{-1}$.   Unitary composition works
as one expects and there is no change to ordinary quantum theory in
regions with unitary evolution.

If the expectation value of an operator $P_\alpha$ is computed on the initial
hypersurface $\sigma_i$, one finds
\begin{equation}
\phantom{\langle}_{(\mu_a)}\langle P_\alpha \rangle=\phantom{\langle}_{(1)}
\langle \psi,i | \mu_a P_\alpha | \psi,i \rangle.
\end{equation}
On the final hypersurface $\sigma_f$, the ``push-forward" of $P_\alpha$ is
$XP_\alpha X^{-1}$ which is assumed here to remain $P_\alpha$.
Its expectation value on $\sigma_f$ is then
\begin{eqnarray}
\phantom{\langle}_{(\mu_b)}\langle XP_\alpha X^{-1} \rangle &=&
\phantom{\langle}_{(\mu_b)}\langle P_\alpha \rangle \\
&=& \phantom{\langle}_{(\mu_b)}\langle \tilde\psi,f| XP_\alpha X^{-1}
| \tilde\psi,f \rangle \nonumber \\
&=& \biggl( \phantom{\langle}_{(1)} \langle \psi,i | X^{\dagger_1} \biggr)
\biggl( (X^{-1})^{\dagger_1} \mu_a X^{-1} \biggr) \biggl( X P_\alpha X^{-1}
\biggr) \biggl(   X |\psi,i \rangle \biggr) \nonumber \\
&=& \phantom{\langle}_{(1)} \langle \psi,i | \mu_a P_\alpha | \psi,i \rangle.
\nonumber
\end{eqnarray}
This agrees with the result computed on $\sigma_i$ and there is no
ambiguity.  As well, the state does not appear in the denominator and
expectation values take their usual linear quantum mechanics form.
Finally, the result does not know of the presence or absence of $X$ and
is not acausally affected by CTC regions.

Turn now to quantum field theory.  To quantize a field, one begins in a
non-interacting initial
region and solves the (initial) free wave operator $L_i$ for its modes, say
$L_i u_j=0$. The modes are normalized in an inner product of the form
(\ref{ip}).  The wave operator is self-adjoint in the measure
density of the inner product.   Using a scalar field as an example, the
quantum field is expanded in terms of the modes with
creation and annihilation operator coefficients
\begin{equation}
\Phi=\sum_j a_j u_j + a_j^{\dagger_a} u_j^*.
\end{equation}
The vacuum is defined as the state annihilated by all the $a_j$,
and the n-particle states are formed by applying a product of
creation operators to the vacuum.

The next step is to compute new mode function for the (final) free
wave operator in a non-interacting final region.
In the original measure density $\mu_a$, this
operator is $L_f$, the usual free wave operator evaluated in the
final region.  It has modes $v_k$.  If a nonunitary evolution $X$
takes place and the measure density in the final region has
changed (\ref{tmd}), then $L_f$ is not the self-adjoint free wave operator
in $\mu_b$, but instead $\tilde L_f= R L_f R^{-1}$ is.  The mode functions
$\tilde v_k$ of $\tilde L_f$ are related to those of $L_f$ by
\begin{equation}
v_k= R^{-1} \tilde v_k.
\end{equation}
This corresponds to Eq. (\ref{trans}) above.

Take the non-interacting case first.  To compute the S-matrix,
the initial modes $u_j$ are evolved into the final region by $X$
where they may then be expanded in terms of the complete set of modes
$\tilde v_k$ (and $\tilde v_k^*$).  Bogoliubov coefficients are
calculated in the usual way using the inner product with measure
density $\mu_b$. The S-matrix is necessarily unitary because the
inner products of states are preserved by the evolution and because
the basis of final mode functions is complete.   Alternatively, if one
chooses to work always in the original familiar
Hilbert space, one simply uses the final mode functions $v_k$, evolves
the initial modes with $U_X=R^{-1}X$ and computes
Bogoliubov coefficients using the inner product with $\mu_a$.

In the interacting case, attention shifts away from evolution of the mode
functions to the detailed correspondence between initial and final
n-particle states.  The evolution operator $X$ can be formally
expressed as
\begin{equation}
X=\sum_{\{\tilde n\},\{m\}} a_{{\tilde n}m}|\tilde n,f\rangle
\phantom{\langle}_{(\mu_a)}\langle m,i|.
\end{equation}
The $\tilde n$ is short-hand for the detailed structure of the
n-particle states associated with the $\tilde v_k$ modes, and $m$
represents m-particle states associated with the $u_j$ modes.
If $X$ is not unitary in ${\cal H}_a$, then
\begin{equation}
\label{smnu}
X X^{\dagger_a}=R^2\ne 1,
\end{equation}
where $R=R^{\dagger_a}$ is self-adjoint in ${\cal H}_a$. Multiplying
this equation
on the right and left by $R^{-1}$, one finds it defines a unitary transition
matrix $U_X=R^{-1}X$.  This matrix is the S-matrix for the
interacting theory and is unitary by construction.  It is non-trivial
if $U_X$ is not the identity.  Unitarity has again been restored by
removing the nonunitary part of the evolution operator.

For completeness, I remark that it follows from Eqs. (\ref{tmd})
and (\ref{smnu}) that
\begin{equation}
X^{\dagger_b} X=X X^{\dagger_a} =R^2 \ne 1.
\end{equation}
Thus, $R$ is self-adjoint in ${\cal H}_b$ as well as ${\cal H}_a$.
Multiplying by $R^{-1}$ on the left and the right, one finds that
$\tilde U_X= X R^{-1}$ is unitary in ${\cal H}_b$.  This explains the
role of the second polar decomposition $X=\tilde U_X R$:  it provides
the unitary operator for evolution which remains in ${\cal H}_b$.

This story can be retold in the decoherence functional formulation.
There, unitarity is restored by appropriate definition of the trace.
The trace in the decoherence functional is
taken over a final hypersurface.  One must make sure that it is the trace
which is physically equivalent to one defined on an initial surface, say
before the CTC region.  This can be done by requiring that the trace of
$\rho_i$ is preserved under the evolution.  On an initial hypersurface
$\sigma_i$, the measure density is $\mu_a$ and the trace is
\begin{equation}
{ \rm Tr}_{(\mu_a)}(\rho_i)= { \rm Tr}_{(1)}(\rho_i\mu_a).
\end{equation}
On the final hypersurface $\sigma_f$, the measure density changes to $\mu_b$
and one takes the trace of the evolved
density matrix $X\rho_i X^{\dagger_b}$
\begin{eqnarray}
{ \rm Tr}_{(\mu_b)}(X\rho_i X^{\dagger_b})&=& { \rm Tr}_{(1)}(X\rho_i
\mu_b X^{\dagger_b}) \\
&=& { \rm Tr}_{(1)}(X\rho_i X^{\dagger_1} \mu_b ). \nonumber
\end{eqnarray}
The traces of $\rho_i$ on $\sigma_i$ and $\sigma_f$ are equal if
\begin{equation}
\label{tmeas}
\mu_b=(X^{-1})^{\dagger_1} \mu_a X^{-1}= \mu_a (X X^{\dagger_a})^{-1}=\mu_a
(X^{\dagger_b} X)^{-1},
\end{equation}
as above.  It is important to emphasize that care must be taken
with the use of the different types of  adjoint as one changes
the hypersurface/Hilbert space where one is taking the trace
and as one adjusts whether the measure density is implicit or explicit.

I remark that the trace of $\rho_i^2$ is also preserved
\begin{eqnarray}
{ \rm Tr}_{(\mu_a)}(\rho_i^2)&=& { \rm Tr}_{(1)}(\rho_i \mu_a \rho_i
\mu_a) \\
&=& { \rm Tr}_{(1)}(X\rho_i X^{\dagger_1} \mu_b X\rho_i X^{\dagger_1}
\mu_b) \nonumber \\
&=& { \rm Tr}_{(\mu_b)}((X\rho_i X^{\dagger_b})^2) .\nonumber
\end{eqnarray}
Thus, this proposal does not provide an interpretation for evolutions
which are not invertible in which a pure state evolves to a mixed state.
Such an evolution would not be described by $X\rho_i X^{\dagger_b}$, but
the point is emphasized so that there should be no mistake.

The decoherence functional for one projection made on the hypersurfaces
$\sigma_1$ and $\sigma_2$ before and after the CTC region is given by
\begin{eqnarray}
D(\alpha',\beta';\alpha,\beta)&=& \\
&&\hspace{-1in}= { \rm Tr}_{(\mu_b)}\biggl(
\tilde P_{\alpha'}^{(2)}(\sigma_2)
XP_{\beta'}^{(1)}(\sigma_1) \rho_i P_{\beta}^{(1)\dagger_b}(\sigma_1)
X^{\dagger_b} \tilde P_{\alpha}^{(2)\dagger_b}(\sigma_2) \biggr)\nonumber\\
&&\hspace{-1in}= { \rm Tr}_{(1)}\biggl( \tilde P_{\alpha'}^{(2)}(\sigma_2) X
P_{\beta'}^{(1)}(\sigma_1)
\rho_i P_{\beta}^{(1)\dagger_1}(\sigma_1) X^{\dagger_1}
\tilde P_{\alpha}^{(2)\dagger_1}(\sigma_2) \mu_b \biggr). \nonumber
\end{eqnarray}
The projection operators are in the Heisenberg picture as defined
in Eq. (\ref{projs}).  This means that $P_\alpha^{(1)}(\sigma_1)$
is self-adjoint in ${\cal H}_a$ on the initial hypersurface while
$\tilde P_\alpha^{(2)}(\sigma_2)$ is self-adjoint in ${\cal H}_b$ on the
final hypersurface.  In the first form, the trace is evaluated on the final
surface in ${\cal H}_b$.  The second form makes the dependence on
the measure density explicit.  This satisfies Hartle's axioms\cite{Har2}
for a decoherence functional.

By using Eq. (\ref{tmeas}), the
decoherence functional can be expressed in terms of a trace in
${\cal H}_a$ on the initial hypersurface,
\begin{eqnarray}
\label{Xdf}
D(\alpha',\beta';\alpha,\beta)&=& \\
&&\hspace{-1in} = { \rm Tr}_{(1)}\biggl( \tilde P_{\alpha'}^{(2)}(\sigma_2)
X P_{\beta'}^{(1)}(\sigma_1)\rho_i P_{\beta}^{(1)\dagger_1}(\sigma_1)
X^{\dagger_1} \tilde P_{\alpha}^{(2)\dagger_1}(\sigma_2) (X^{-1})^{\dagger_1}
\mu_a X^{-1} \biggr).\nonumber \\
&&\hspace{-1in}= { \rm Tr}_{(\mu_a)}\biggl( X^{-1} \tilde
P_{\alpha'}^{(2)}(\sigma_2)
XP_{\beta'}^{(1)}(\sigma_1) \rho_i P_{\beta}^{(1)\dagger_a}(\sigma_1) (X^{-1}
\tilde P_{\alpha}^{(2)}(\sigma_2) X)^{\dagger_a} \biggr).\nonumber
\end{eqnarray}
The effect of the change in measure has been to pull back the projection
operator $\tilde P_\alpha^{(2)}(\sigma_2)$ to the initial hypersurface by
the adjoint action of $X^{-1}$.  Since $(X^{-1})^{\dagger_1}
\mu_a X= \mu_a R^{-2}$, the decoherence functional  can be reexpressed as
a trace in ${\cal H}_a$ on the final hypersurface
\begin{eqnarray}
D(\alpha',\beta';\alpha,\beta)&=& \\
&&\hspace{-1in} = { \rm Tr}_{(1)}\biggl( \tilde P_{\alpha'}^{(2)}(\sigma_2)
X P_{\beta'}^{(1)}(\sigma_1)\rho_i P_{\beta}^{(1)\dagger_1}(\sigma_1)
X^{\dagger_1} \tilde P_{\alpha}^{(2)\dagger_1}(\sigma_2)  \mu_a R^{-2}
\biggr).\nonumber \\
&&\hspace{-1in}= { \rm Tr}_{(\mu_a)}\biggl(
\tilde P_{\alpha'}^{(2)}(\sigma_2) XP_{\beta'}^{(1)}(\sigma_1) \rho_i
P_{\beta}^{(1)\dagger_a}(\sigma_1) X^{\dagger_a}
\tilde P_{\alpha}^{(2)\dagger_a}(\sigma_2) R^{-2} \biggr).\nonumber
\end{eqnarray}

There is a factor $R^{-2}$ present here that is not in the naive
decoherence functional, cf. the numerator of Eq. (\ref{ndf}).
A more subtle but equally important difference is that the
projection operator after the CTC region is $\tilde P_\alpha^{(2)}(\sigma_2)$
which is self-adjoint in ${\cal H}_b$, not ${\cal H}_a$!
Using Eq. (\ref{optrans}) to transform to the projection
operator  $P_\alpha^{(2)}(\sigma_2)$ which is self-adjoint in ${\cal H}_a$,
the expression for the decoherence functional is found
\begin{equation}
D(\alpha',\beta';\alpha,\beta)=
{ \rm Tr}_{(\mu_a)}\biggl( \tilde P_{\alpha'}^{(2)}(\sigma_2)
U_X P_{\beta'}^{(1)}(\sigma_1) \rho_i P_{\beta}^{(1)\dagger_a}(\sigma_1)
U_X^{\dagger_a} P_{\alpha}^{(2)\dagger_a}(\sigma_2) \biggr).
\end{equation}
This clearly shows that the mechanism for unitarity restoration is
to replace the nonunitary evolution $X$ by its unitary part $U_X$.

Finally, the trace in the decoherence functional can be shifted to the
initial hypersurface by inserting $U_X U_X^{-1}=1$ at the end of the trace.
Using cyclicity, one sees that the decoherence functional is built from
the Heisenberg evolved projection operators (\ref{Hop})
\begin{eqnarray}
D(\alpha',\beta';\alpha,\beta)&=& \\
&&\hspace{-1in} =  { \rm Tr}_{(\mu_a)}\biggl( U_X^{-1}
P_{\alpha'}^{(2)}(\sigma_2) U_X P_{\beta'}^{(1)}(\sigma_1) \rho_i
P_{\beta}^{(1)\dagger_a}(\sigma_1) (U_X^{-1}
\tilde P_{\alpha}^{(2)}(\sigma_2) U_X)^{\dagger_a} \biggr).\nonumber
\end{eqnarray}

Working with Eq. (\ref{Xdf}), one can consider a few cases to discover
the dependence of the decoherence functional on $X$.  Suppose first that
there is only a projection before the CTC region so that
$\tilde P_\alpha^{(2)}(\sigma_2)$ and
$\tilde P_{\alpha'}^{(2)}(\sigma_2)$ are not
present.  Then, the $X$ factors cancel and there is no dependence on $X$.
A CTC region in the future does not acausally affect prediction made in the
present. Alternatively, suppose that $\tilde P_\alpha^{(2)}(\sigma_2)$ and
$\tilde P_{\alpha'}^{(2)}(\sigma_2)$ are in a region spacelike separated
from the CTC region so that they commute with $X$.  Again the $X$ factors
cancel and the CTC region has no acausal effect.  Only if
$\tilde P_\alpha^{(2)}(\sigma_2)$ does not commute with $X$ will there
be an effect.  This presumably can happen if
$\tilde P_\alpha^{(2)}(\sigma_2)$ is in the causal future of the CTC region.

One can also compute probabilities for alternatives in the CTC
region by inserting projection operators between periods of
nonunitary evolution.  These alternatives must of course be members
of an exhaustive and exclusive set.  Because there is not a spacelike
hypersurface which intersects the CTC region, it is not clear how to describe
such a computation in terms of states at the instant projection is
made.  The computation can be made in the Heisenberg picture by
pulling the projection operator back to the initial hypersurface and
computing the expectation value there, as in Eq. (\ref{Xdf}).

It should be mentioned that one cannot necessarily detect CTC regions in
one's past because experimentally we only have access to an
effective initial density
matrix.  Our effective initial density matrix may have been renormalized
by $X$ from the true initial density matrix, but we would not be able to
detect this without knowing the initial density matrix.  On the other hand,
if one has a theory for the initial
condition of the universe, as in the Hartle-Hawking no-boundary proposal,
predictions based on this initial condition could differ depending on
the presence or absence of
CTC regions and this might be detectable.   Otherwise, the best means of
detecting a CTC region is to measure a change in amplitudes as
one evolves past it--in this way, one could compare results
before and after and see a possible difference due to the
noncommutativity with $X$.  Sadly, there seems little likelihood of an
experimental test of this kind.

For completeness, one can consider a time-symmetric decoherence
functional in which a final state density matrix is also imposed.  The
objective here is to verify the time symmetry of the decoherence functional
in the presence of nonunitary evolution.  The subtlety in this is again
taking care with the hypersurface on which the trace is evaluated. To be
completely explicit, consider a situation in which an evolution $X$ from
an initial surface $\sigma_a$ to an intermediate surface $\sigma_1$ is
followed by a projection $P_\alpha^{(1)}$ and then by subsequent evolution $Y$
to a final surface $\sigma_b$.  (Note that in contrast to the discussion
above, the projection operator is in the Schr\"odinger picture and the
intermediate evolutions are included explicitly as $X$ and $Y$.)
Neither $X$ nor $Y$ will be assumed to be unitary.  To be clear about the
definition of adjoints, let $\mu_a$ be the (coordinate-valued) measure
density on the initial surface $\sigma_a$.  The adjoint $\dagger_a$ on
$\sigma_a$ is  defined by $\mu_a X^{\dagger_a}= X^{\dagger_1} \mu_a$ and
the adjoint $\dagger_b$ is defined similarly on the final surface using
the measure density $\mu_b$. The projection is self-adjoint
$P_\alpha^{(1)}=P_\alpha^{(1) \dagger_{(1)}}$ on the intermediate surface
where it is defined, using the induced measure density $\mu_{(1)}$
there\cite{sadj}.

On the final hypersurface the decoherence functional is
\begin{eqnarray}
\label{tsdf}
D(\alpha';\alpha)&=&{{ \rm Tr}_{(\mu_b)}\biggl( \rho_f Y P_{\alpha'}^{(1)}
X \rho_i X^{\dagger_b} P_{\alpha}^{(1)\dagger_b} Y^{\dagger_b} \biggr) \over
{ \rm Tr}_{(\mu_b)} \biggl( \rho_f YX \rho_i X^{\dagger_b} Y^{\dagger_b}
\biggr)} \\
&=&{{ \rm Tr}_{(1)}\biggl( \rho_f \mu_b YP_{\alpha'}^{(1)} X\rho_i
X^{\dagger_1} P_{\alpha}^{(1)\dagger_1} Y^{\dagger_1} \mu_b\biggr) \over
{ \rm Tr}_{(1)} \biggl( \rho_f \mu_b YX \rho_i X^{\dagger_1}
Y^{\dagger_1}\mu_b \biggr)}. \nonumber
\end{eqnarray}
The final density matrix acquires a factor of the measure density
when one introduces the measure explicitly.  This is easily
confirmed with a calculation in terms of states where a trace over two
density matrices leads to a product of two inner products and hence
requires two measure density factors.  Using the transformation formula
for the measure density,
$\mu_b=(YX)^{-1\,\dagger_1} \mu_a (YX)^{-1}$, the denominator is seen
to be symmetric
\begin{equation}
{ \rm Tr}_{(1)} \biggl( \rho_f \mu_b YX \rho_i X^{\dagger_1} Y^{\dagger_1}
\mu_b \biggr) = { \rm Tr}_{(1)} \biggl(  \rho_i \mu_a X^{-1}Y^{-1}
\rho_f (Y^{-1})^{\dagger_1} (X^{-1})^{\dagger_1} \mu_a \biggr).
\end{equation}

The time-symmetry of the numerator is slightly more difficult to establish
due to subtlety with the
adjoints.  Two side results are needed.  First, the inverse and
adjoint ($\dagger_1$) operations commute
\begin{equation}
(X^{-1})^{\dagger_1}=(X^{\dagger_1})^{-1},
\end{equation}
which follows by taking the adjoint of $X X^{-1}=1$.  Next,
self-adjointness of $P_\alpha^{(1)}$ on the intermediate surface implies
self-adjointness of the projection pulled back to the initial surface
$X^{-1}P_\alpha^{(1)} X$ when the measure density transforms appropriately
for nonunitary $X$, cf. Eq. (\ref{sadja}).
Note however that $P_\alpha^{(1)}\ne P_\alpha^{(1)\dagger_a}$ if $X$ is
not unitary.

In the numerator of (\ref{tsdf}), consider the factor $\mu_b Y
P_{\alpha'}^{(1)}
X$.    This can be reexpressed with $\mu_a$ on the right on transforming
the measure and using the computation (\ref{sadja})
\begin{eqnarray}
\mu_b YX(X^{-1} P_{\alpha'}^{(1)}X)&=&
((YX)^{-1})^{\dagger_1}\mu_a \mu_a^{-1} X^{\dagger_1 }
P_{\alpha'}^{(1)\dagger_1} (X^{-1})^{\dagger_1} \mu_a \nonumber \\
&=&( Y^{-1})^{\dagger_1} P_{\alpha'}^{(1)\dagger_1} (X^{-1})^{\dagger_1} \mu_a.
\end{eqnarray}
Similarly, one has
\begin{equation}
X^{\dagger_1} P_{\alpha}^{(1)\dagger_1}Y^{\dagger_1} \mu_b=
\mu_a X^{-1} P_\alpha^{(1)} Y^{-1}.
\end{equation}
These can be combined to show that the numerator of the final equality in
(\ref{tsdf}) is
\begin{eqnarray}
{ \rm Tr}_{(1)}\biggl( \rho_f ( Y^{-1})^{\dagger_1} P_{\alpha'}^{(1)\dagger_1}
(X^{-1})^{\dagger_1} \mu_a
\rho_i \mu_a X^{-1} P_\alpha^{(1)} Y^{-1} \biggr)  &=& \\
&&\hspace{-2in} = { \rm Tr}_{(\mu_a)} \biggl( \rho_i X^{-1}
P_\alpha^{(1)} Y^{-1}
\rho_f (Y^{-1})^{\dagger_a} P_\alpha^{(1)\dagger_a} (X^{-1})^{\dagger_a}
\biggr). \nonumber
\end{eqnarray}
This is the time-symmetric form of the numerator, and one concludes that
the decoherence functional with initial and final density matrices
is time-symmetric.

To summarize, it has been shown that evolution by an invertible
``nonunitary'' operator $X$ with $X X^{\dagger_a}=R^2\ne 1$ can be made
unitary by transforming the measure density on the final hypersurface
appropriately, Eq. (\ref{tmd}). Equivalently, one can stay in the original
Hilbert space by removing the nonunitary part
of $X$ to obtain the unitary evolution operator $U_X=R^{-1}X$.
Both procedures lead to unambiguous results for the
expectation values of observables. The results are causal and do not involve
nonlinear modifications of quantum mechanics.
This proposal for handling nonunitary evolution can be stated in the
form of a decoherence functional in the language of Gell-Mann and Hartle
generalized quantum mechanics.  The decoherence functional with initial
and final density matrices imposed as conditions is time-symmetric.

The conclusion is that one can consistently make quantum mechanical
computations in spacetimes containing closed timelike curves without
loss of probability or causality if evolution is described by an
invertible nonunitary operator.  A region of closed timelike curves
could only be detected if one interacted with something in its causal
future.

Acknowledgements.  I would like to thank D. Craig, D. Deutsch, J. Hartle, D.
Politzer, S. Rosenberg, J. Whelan and the other participants of the workshop
on closed timelike curves at the Isaac Newton Institute for many
stimulating discussions, and C. Fewster, J. Friedman and
T. Jacobson for further helpful conversations.


\begin{thebibliography}{99}
\bibitem{sadj}  The projection is assumed to be defined on a spacelike
hypersurface.  In the broader context of generalized quantum mechanics
where projections may be onto regions of spacetime, greater care must be
taken in defining the adjoint.
\end{thebibliography}
\end{document}